\documentclass[conf]{new-aiaa}
\usepackage[utf8]{inputenc}
\usepackage{mathtools}
\usepackage{graphicx}
\usepackage{amsmath}
\usepackage[version=4]{mhchem}
\usepackage{siunitx}
\usepackage{booktabs}
\usepackage{longtable,tabularx}
\usepackage{todonotes}
\usepackage{caption}
\usepackage{subcaption}

\title{Adaptive Gain Scheduling using Reinforcement Learning for Quadcopter Control}

\author{Mike Timmerman}
\affil{Dept. of Aeronautics and Astronautics, Stanford University}
\author{Aryan Patel}
\affil{Dept. of Mechanical Engineering, Stanford University}
\author{Tim Reinhart}
\affil{Dept. of Computer Science, Stanford University}

\begin{document}

\maketitle

\begin{abstract}
The paper presents a technique using reinforcement learning (RL) to adapt the control gains of a quadcopter controller. Specifically, we employed Proximal Policy Optimization (PPO) to train a policy which adapts the gains of a cascaded feedback controller in-flight. The primary goal of this controller is to minimize tracking error while following a specified trajectory. The paper's key objective is to analyze the effectiveness of the adaptive gain policy and compare it to the performance of a static gain control algorithm, where the \textit{Integral Squared Error} and \textit{Integral Time Squared Error} are used as metrics. The results show that the adaptive gain scheme achieves over 40$\%$ decrease in tracking error as compared to the static gain controller.
\end{abstract}

\section{Introduction}
In recent years, Reinforcement Learning algorithms have been applied to a variety of fields, including robotics, trajectory planning, chess games, and video games. This paper aims to apply reinforcement learning techniques to train the gains for a quadcopter controller. A quadcopter is an autonomous drone which achieves flight and thrust through four separate rotors. These rotors are separately controlled using a controller which enables the quadcopter to closely follow a planned trajectory. 

As the dynamics of a quadcopter are inherently unstable, unlike planes the controller needs to act quickly based on information provided by sensors to maintain the planned trajectory \cite{final62}. These sensors typically include an accelerometer and gyroscope. To ensure that the quadcopter follows a given trajectory, controllers are implemented. This project plans to utilize RL algorithms to successfully optimize the gain values for a given cascaded feedback controller. Many RL controllers need to go through a training period where the controller behaves poorly. To avoid damage to an actual quadcopter due to loss of control during the training phase, a virtual environment was setup that simulated the dynamics. To avoid computational complexity, the quadcopter is confined to a 2-dimensional space, with three degrees of freedom: translation in the $x$- and $y$-axes and rotation in the $x-y$-plane. For each time step in the simulation, the motion of the quadcopter is calculated using it's dynamics model. 

In this paper, we attempt to provide a successful implementation of the Proximal Policy Optimization (PPO) RL algorithm, which is used to generate a policy to adapt the gains for a cascaded feedback controller. The paper first provides a brief survey of related literature about implementing RL algorithms to classic control techniques. It then provides a walk through of the virtual environment as well as an overview of the PPO algorithm. The results are then presented comparing the effect of RL-controlled trajectory to classic control schemes. Finally, conclusions and possible future work are presented.

\section{Related Work}
% Quadcopter controllers are generally implemented using variants of Proportional-Integral-Derivative (PID) controllers. These controllers perform well because the dynamics of a quadcopter are known, so fine tuning the model for appropriate gains is possible. Static PID controllers have fast convergence and tuned output. However, static PID controllers may not perform well when adapting to new environments.  To solve this, variants of the PID controller have been used.

Learning based techniques have been utilized either to supplement or directly replace controllers. Wang et al. (2007) proposed an actor-critic method for tuning a PID controller and demonstrated its performance on a simulated non-linear system. The actor and critic made use of a neural network with a single hidden radial basis function layer \cite{Wang_RL}. Wang et al. showed that the PID tuned by this method out-performed a PID tuned using the Zeigler-Nichols method. Arzaghi-Haris (2008) subsequently used this reinforcement learning method and network architecture to tune a PID to regulate wind turbine rotor speed to the optimal speed in order to maximise the power generated \cite{Arzaghi_Harris}. Shuprajhaa et al. (2022) used a modified PPO algorithm to tune a PID controller for open loop unstable processes \cite{Shiva}.  In the scope of our class, Shipman and Coetzee used the A2C algorithm to tune a PI controller for process control \cite{RLPI_Tune}.  Fang-I Hsiao and Cheng-Min Chiang used A2C and PPO directly in place of a static PID controller to study the stability of a hovering quadcopter \cite{final62}. In this study, we aim to compare the results of using RL-learning to tune the gains of a cascade PD controller.

\section{Environment}
The environment contains the representation of the $\textit{Markov Decision Process (MDP)}$. It is implemented using the $\textit{Gymnasium}$ API. The environment, describing the MDP, consists of several key components:
\begin{enumerate}
    \item Agent: The agent is the defined as the drone-controller combination which interacts with the environment, providing observations to the RL policy and receiving actions from the policy. The agent interacts with the environment by tracking pre-generated trajectories.
    \item Transition: Transitions are based on the dynamics and architecture of the base controller. The drone dynamics used are the planar quadcopter dynamics consisting of 6 states, given in \autoref{eq:EOM}.
    \begin{equation}
        \label{eq:EOM}
        \begin{bmatrix}
            \Dot{p_x} \\
            \Dot{p_y} \\
            \Dot{\theta} \\
            \Dot{V_x} \\
            \Dot{V_y} \\
            \Dot{\omega} \\
        \end{bmatrix} = 
        \begin{bmatrix}
            V_x \\
            V_y \\
            \omega \\
            -\frac{(T_1 + T_2)sin(\theta) - C_{d_V} V_x}{m} \\
            \frac{(T_1 + T_2)cos(\theta) - C_{d_V} V_y}{m} - g \\
            \frac{(T_2 - T_1)l - C_{d_{\omega}} \omega}{m} \\
        \end{bmatrix}
    \end{equation}
    The base controller architecture is given in \autoref{fig:base_controller}. The architecture is a cascaded feedback structure with proportional gains in each feedback loop \cite{handbook}. The agents' parameter explanations and values are given in \autoref{tab:params}.
    \begin{figure}[h!]
        \centering
        \includegraphics[width=\textwidth]{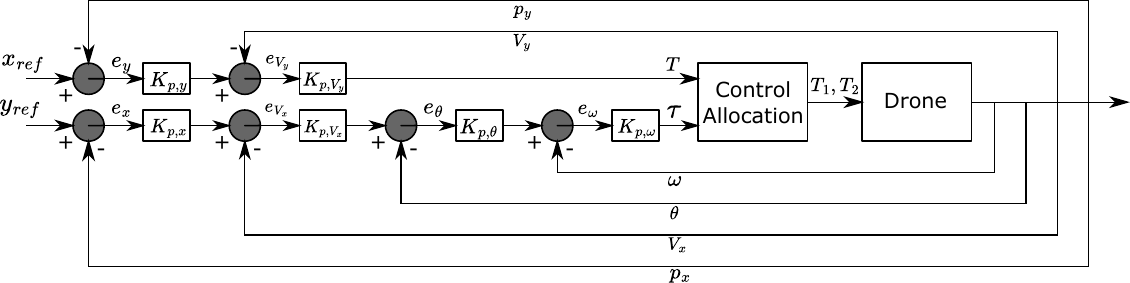}
        \caption{Base controller architecture}
        \label{fig:base_controller}
    \end{figure}%
    \begin{table}[h!]
        \centering
        \caption{Agent parameters}
        \label{tab:params}
        \resizebox{0.5\columnwidth}{!}{%
        \begin{tabular}{c|l|c}
        \toprule
        Parameter & Explanation & Value \\
        \midrule
        m & mass & $2.5 [kg]$ \\
        I & inertia & $1.0 [kgm^3]$ \\
        l & length of propeller to center of mass & $1.0 [m]$ \\
        g & gravitational constant & $9.807 [\frac{m}{s^2}]$ \\
        $C_{d_v}$ & drag coefficient for to linear velocity & $0.25 [-]$ \\
        $C_{d_{\omega}}$ & drag coefficient for to rotational velocity & $0.02255 [-]$ \\
        \bottomrule
        \end{tabular}}
    \end{table}%
\newpage
    \item Action Space:
        The action space is a subset of $\mathbb{R}^6$, where each element is bounded between $[-1, 1]$. Each element corresponds to a proportional gain in the base controller, given in \autoref{fig:base_controller}. The actions are rescaled to a set range, defined for each gain, as given in \autoref{tab:gain_ranges}. These regions were established based on a frequency-domain analysis.
        \begin{table}[h!]
            \centering
            \caption{Base controller proportional gain ranges}
            \label{tab:gain_ranges}
            \begin{tabular}{l|c|c|c|c|c|c}
            \toprule
                & $K_{p_x}$ & $K_{p_{V_x}}$ & $K_{p_{\theta}}$ & $K_{p_{\omega}}$ &  $K_{p_y}$ & $K_{p_{V_y}}$ \\
            \midrule
                Range & $[0.5, 2.0]$ & $[-0.5, -0.1]$ & $[5.0, 10.0]$ & $[10.0, 16.0]$ & $[0.5, 3.0]$ & $[5.0, 15.0]$ \\
            \bottomrule
            \end{tabular}
        \end{table}
    \item State Space:
        The state space is $\mathbb{R}^6$. Each element corresponds to an error term in the base controller, i.e. the observations are the vector $[e_x; e_{V_x}; e_{\theta}; e_{\omega}; e_{y}; e_{V_y}]$.
    \item Reward: The reward function consists of four parts corresponding to four distinct scenarios in which the drone might end up: 
    \begin{enumerate}
        \item Time-out: the drone took longer than $20\%$ of the expected time to reach to target position
        \item Deviation: the drone deviated more than $10[m]$ from the current reference position
        \item Success: the drone reached its target successfully
        \item No termination: none of the above conditions are met
    \end{enumerate}
    The reward values for each of these cases is given by the reward function given in \autoref{eq:reward_function}:
    \begin{equation}
        \label{eq:reward_function}
        R(s, a) = 
        \begin{cases}
            -1 & \text{for time-out} \\
            -5 & \text{for deviation} \\
            \frac{10}{ISE} & \text{for success} \\
            0.05\min\left(\frac{\text{previous deviation}}{\text{current deviation}}-1, 2\right)& \text{for no termination} \\
        \end{cases}
    \end{equation}
    This reward function punishes the policy when it does not reach the target and rewards it when it does, giving a higher reward for a smaller integral squared error (ISE). When no termination occurs, a positive reward is given when the deviation from the reference is decreased compared to the previous deviation. The reward then falls in the range $[-0.05, 0.1]$.
\end{enumerate}

\section{Method \\}

To develop a policy for optimizing the gains seen by the cascade PD controller, the Proximal Policy Optimization (PPO) method was chosen. Both these methods are policy gradient methods that optimize the policy based on calculating the policy gradient \cite{AA228_text}. The policy gradient of a policy $\pi$ parameterized by $\theta$ can be calculated using \autoref{eq:pol_grad} \cite{AA228_text}:
    \begin{equation}
        \label{eq:pol_grad}
        \nabla_{\theta} J(\theta) = \mathbb{E}_{\pi} [Q(s,a) \nabla_{\theta} \log \pi(a|s)]
    \end{equation}
Then, gradient descent is used to update the policy.

\subsection{Proximal Policy Optimization (PPO)}
Proximal Policy Optimization (PPO) is the chosen learning method to learn the optimal policy. PPO is a reinforcement learning algorithm based on policy gradient methods. However, PPO differs from standard policy gradient methods, which are on-policy. This means that they require re-sampling after each update to evaluate the new policy. This method is inefficient due to time consumption in interacting with the environment. In contrast PPO allows multiple updates to the gradient by utilizing different batches of the sampled data. Hence, the training speed is faster.

The Proximal Policy Optimization algorithm combines ideas from A2C (an actor-critic method) and TRPO (Trust Region Policy Optimization) in that it has multiple workers and uses a trust region to improve the actor. PPO works by optimizing a "surrogate" objective function using stochastic gradient ascent \cite{PPO}. In addition, PPO allows for multiple epochs of minibatch updates using a novel objective function as shown by Schulman et al. \cite{PPO}.  \\

\subsection{Training Set-up}
The RL policy is required to learn how to tune the gains such that it enables the drone to follow a reference trajectory. In order to achieve this goal, a training environment is set up in which the drone is set to track a step reference in x- and y-position of $1 [m]$.

A total of $\num{2.4e5}$ steps are made, which corresponds to a timestep in the simulation. $3$ environments are run in parallel, each having their own agent taking steps in the environment. On each policy update iteration, $6144$ steps are used to make the update, hence $2048$ steps in each environment. The gradient descent update rule is run with a batch size of $64$. The discount factor is set at $\gamma = 0.99$ and the learning rate at $\eta=\num{3e-4}$.

\section{Results}
\subsection{Training}
\begin{figure}[h!]
    \centering
    \includegraphics[width=0.9\textwidth]{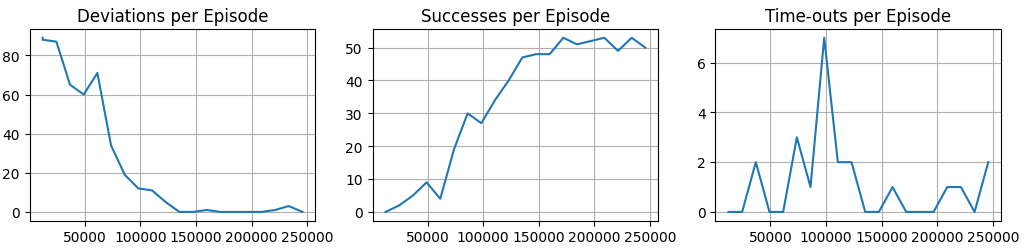}
    \caption{Success rate monitored throughout the training process}
    \label{fig:success_rate}
\end{figure}

The training progress is monitored every two iterations, hence one monitoring episode includes $12288$ steps. \autoref{fig:success_rate} shows three plots in which the number of successes, deviations and time-outs are shown per episode. A clear trend is noticeable where the number of deviations and time-outs decreases as the training progresses, while the number of successes increases.

\begin{figure}[h!]
    \centering
    \includegraphics[width=0.9\textwidth]{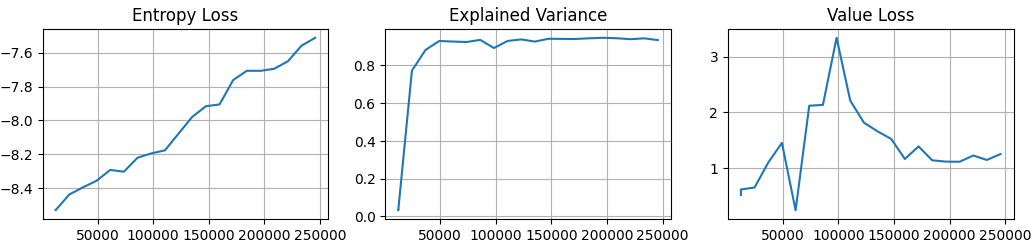}
    \caption{Values monitoring training progress}
    \label{fig:training_progress}
\end{figure}

During the training process multiple values measuring the learning progress are logged. \autoref{fig:training_progress} shows the entropy loss, explained variance, and the value loss. The entropy shows a gradually decreasing trend in magnitude, indicating that the actions become decreasingly random. The explained variance converges to a value of $0.94$, which indicates that the value function is a good predictor of the returns. Lastly, the value loss first increases, indicating that the agent is exploring, after which it decreases, indicating that the reward is stabilizing, eventually converging to a value of $1.15$.

\begin{figure}[h!]
     \centering
     \begin{subfigure}[b]{0.33\textwidth}
         \centering
        \includegraphics[width=\textwidth]{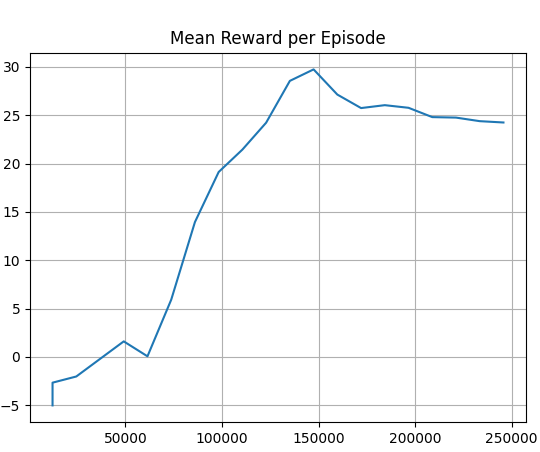}
        \caption{Mean reward accumulated at each episode}
        \label{fig:mean_reward}
     \end{subfigure}
     \hfill
     \begin{subfigure}[b]{0.33\textwidth}
         \centering
         \includegraphics[width=\textwidth]{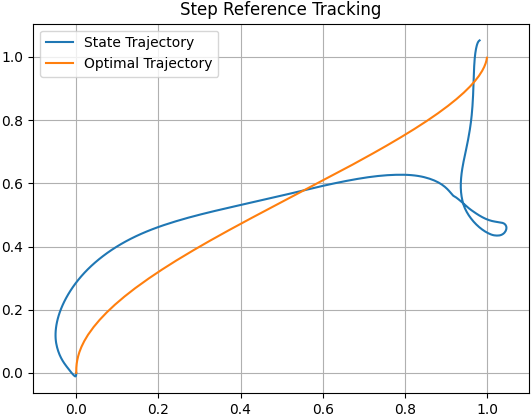}
         \caption{Step tracking performance after 12 iterations}
         \label{fig:low_iter_perf}
     \end{subfigure}
     \hfill
     \begin{subfigure}[b]{0.33\textwidth}
         \centering
         \includegraphics[width=\textwidth]{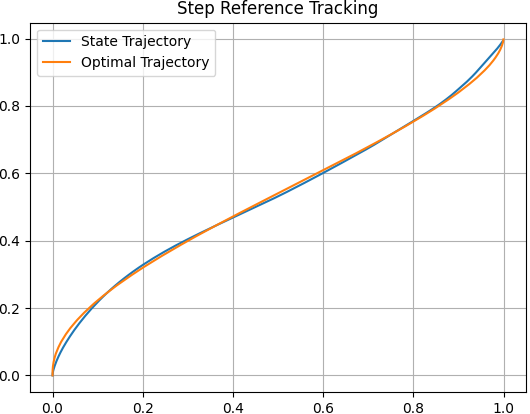}
         \caption{Step tracking performance after 28 iterations}
         \label{fig:high_iter_perf}
     \end{subfigure}
        \caption{Progression in performance throughout training}
        \label{fig:train_performance}
\end{figure}

The RL policy is trained as to collect more rewards as the training progresses. The performance of the policy over the training is shown in \autoref{fig:train_performance}. The mean reward in each episode is shown in \autoref{fig:mean_reward}. As the training progresses, the policy accumulates a higher reward, indicating that the it's performance is improving, up until the point at which it converges, around $\num{1.6e4}$ steps to a value of 25. \autoref{fig:low_iter_perf} shows a trajectory after 12 iterations are completed. Clearly, the policy does not manage to create an efficient trajectory to reach the target. \autoref{fig:high_iter_perf} shows a trajectory after 28 iterations. At this point, the policy manages to nearly recreate an optimized trajectory. The optimized trajectory is obtained through a constraint optimization problem in which a dynamically feasible shortest path trajectory is found. This trajectory is used as comparison for the state trajectory and is not used in the policy training.

\subsection{Evaluation}
The trained policy is evaluated on the performance of tracking a set of three generated trajectories. The performance is measured in terms of the \textit{Integral Squared Error} (ISE) and the \textit{Integral Time Squared Error} (ITSE) and compared to a baseline controller. The baseline controller uses the base controller architecture in \autoref{fig:base_controller} and is manually tuned using loopshaping. The results are shown in \autoref{tab:eval_tab}. Additionally, the percentage decrease of the RL controller compared to the baseline controller is given in the last two columns. From these values, it can be concluded that the RL controller gains an increase in tracking performance of roughly $44\%$.
\begin{table}[h]
    \centering
    \caption{Evaluation metrics}
    \label{tab:eval_tab}
    \resizebox{0.8\columnwidth}{!}{
    \begin{tabular}{c|c|c|c|c|c|c}
        \cline{2-7}
            \multicolumn{1}{c}{} & \multicolumn{2}{c|}{ISE} & \multicolumn{2}{c|}{ITSE} & \multicolumn{2}{c}{Percentage Difference} \\
             \multicolumn{1}{c}{}  & Baseline Controller & RL Controller & Baseline Controller & RL Controller & ISE & ITSE \\
        \midrule
             Trajectory 1 & 0.582 & 0.330 & 71.651 & 37.551 & $-43.3\%$ & $-47.6\%$ \\
             Trajectory 2 & 0.526 & 0.290 & 86.409 & 44.137 & $-44.9\%$ & $-48.9\%$ \\
              Trajectory 3 & 0.357 & 0.205 & 66.681 & 36.120 & $-42.6\%$ & $-45.8\%$ \\
         \bottomrule
    \end{tabular}}
\end{table}

\begin{figure}[h!]
     \centering
     \begin{subfigure}[b]{0.48\textwidth}
         \centering
        \includegraphics[width=\textwidth]{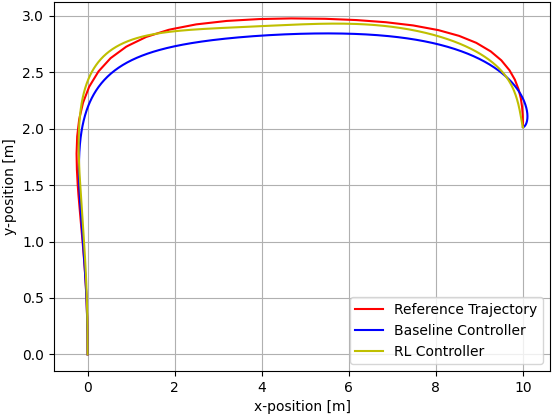}
        % \caption{Mean reward accumulated at each episode}
        % \label{fig:mean_reward}
     \end{subfigure}
     \hfill
     \begin{subfigure}[b]{0.48\textwidth}
         \centering
         \includegraphics[width=\textwidth]{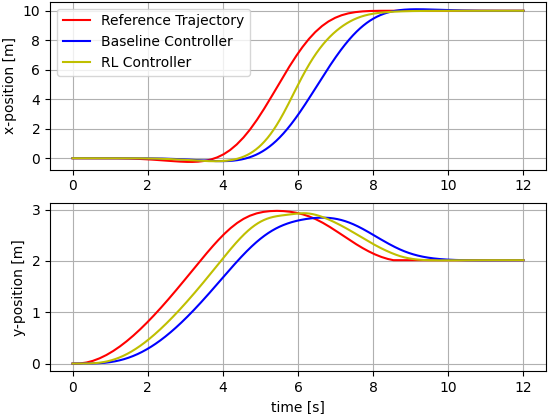}
         % \caption{Step tracking performance after 12 iterations}
         % \label{fig:low_iter_perf}
     \end{subfigure}
        \caption{Comparison of state trajectories}
        \label{fig:evaluation_trajectories}
\end{figure}

\begin{figure}[h]
    \centering
    \includegraphics[width=\textwidth]{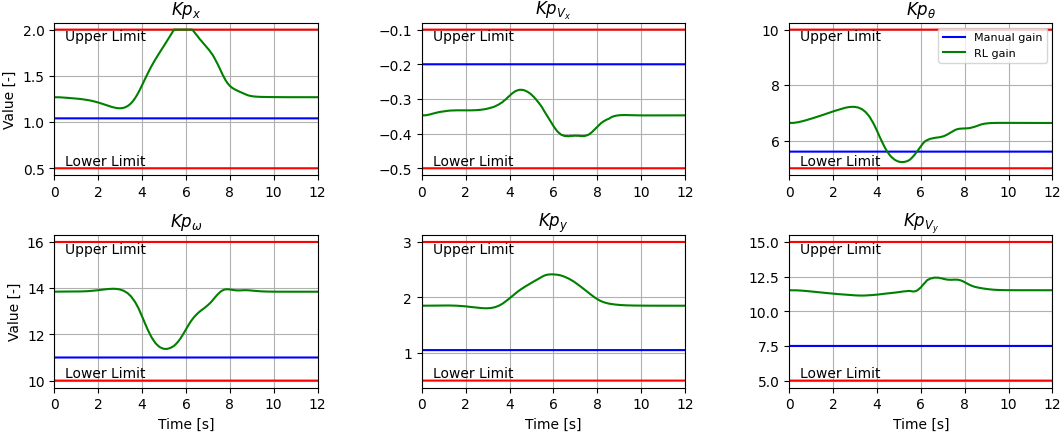}
    \caption{Comparison of controller gains over time}
    \label{fig:gain_trajectories}
\end{figure}

\autoref{fig:evaluation_trajectories} shows a comparison of the state trajectories between the two controllers, as well as the reference trajectory. Clearly, the RL controller is able to track the reference better than the baseline controller. Additionally, \autoref{fig:gain_trajectories} shows a comparison between the manually tuned gains, and the gains selected by the RL policy. From this, it can be seen that the RL controller is able to adapt the gains in response to changes in the reference trajectory.

\section{Conclusion}
This paper investigated the application of a gain scheduled feedback controller in which the gain scheduling is performed using a reinforcement learning policy. The performance of the adaptive controller is compared to a static gain controller with the same architecture. From evaluation results, it can be concluded that the adaptive scheme achieves over a $40\%$ increase in tracking performance compared to the static scheme according to the $\textit{Integral Squared Error}$ and $\textit{Integral Time Squared Error}$ metrics. Future work might explore expanding the results to a 6 degrees of freedom quadcopter. Additionally, stability guarantees when using this adaptive scheme should be investigated. Lastly, testing the scheme on a real drone to evaluate its performance when dealing with real-life limitations and uncertainty.

% \section{Contribution}

% \textbf{Mike Timmerman:} Set up baseline controller and learning implementation using PPO, wiritng of report. 

% \textbf{Aryan Patel:} Helped setup RL method for controller. Wrote major section of report.

% \textbf{Tim Reinhart:} Implementing and setting up of the environment using gymansium, evaluation of results of the trained RL policy, writing of report. 

% \section*{Appendix}

% An Appendix, if needed, should appear before the acknowledgments.

\section{Acknowledgements}
This project was completed as part of the Stanford University course on Decision Making under Uncertainty; AA228.

\bibliography{sample}

\end{document}